\documentclass{article}
\usepackage{arxiv}
\usepackage[utf8]{inputenc} 
\usepackage[T1]{fontenc}    
\usepackage{hyperref}       
\usepackage{url}            
\usepackage{booktabs}       
\usepackage{amsfonts}       
\usepackage{nicefrac}       
\usepackage{microtype}      
\usepackage{lipsum}
\usepackage{graphicx}
\usepackage[style=numeric,sorting=none]{biblatex}
\usepackage{hyperref}
\hypersetup{
    colorlinks=true,
    linkcolor=blue,
    filecolor=magenta,      
    urlcolor=cyan,
}
\usepackage{caption}
\title{Assessing the use of transaction and location based insights derived from Automatic Teller Machines (ATM's) as near real time “sensing” systems of economic shocks}
\author{
  Dharani Dhar Burra\\
  United Nations Global Pulse\\
  Pulse Lab Jakarta, Indonesia\\
  \texttt{dharani.burra@un.or.id} \\
  \And
  Sriganesh Lokanathan\\
  United Nations Global Pulse\\
  Pulse Lab Jakarta, Indonesia\\
  \texttt{sriganesh.lokanathan@un.or.id}\\
}
\begin{document}
\maketitle
\begin{abstract}
 Big data sources provide a significant opportunity for governments and development stakeholders to “sense” and identify in near real time, economic impacts of shocks on populations at high spatial and temporal resolutions. In this study, we assess the potential of transaction and location based measures obtained from automatic teller machine (ATM) terminals, belonging to a major private sector bank in Indonesia, to sense in near real time, the impacts of shocks across income groups. For each customer and separately for years 2014 and 2015, we model the relationship between aggregate measures of cash withdrawals for each year, total inter-terminal distance traversed by the customer for the specific year and reported customer income group. Results reveal that the model was able to predict the corresponding income groups with 80\% accuracy, with high precision and recall values in comparison to the baseline model, across both the years. Shapley values suggest that the total inter-terminal distance traversed by a customer in each year differed significantly between customer income groups. Kruskal-Wallis test further showed that customers in the lower-middle class income group, have significantly high median values of inter-terminal distances traversed (7.21 Kms for 2014 and 2015) in comparison to high (2.55 Kms and 0.66 Kms for years 2014 and 2015), and low (6.47 Kms for 2014 and 2015) income groups. Although no major shocks were noted in 2014 and 2015, our results show that lower-middle class income group customers, exhibit relatively high mobility in comparison to customers in low and high income groups. Additional work is needed to leverage the sensing capabilities of this data to provide insights on, who, where and by how much is the population impacted by a shock to facilitate targeted responses.
\end{abstract}
\section{Introduction}
Big (or Alternative) data sources provide an opportunity for use as early warning and “sensing” systems, to detect economic impacts of shocks on populations in near real time and at across large spatial scales. Such high-resolution insights are necessary for development agencies to plan, implement and also monitor the impacts of their interventions, specifically in response to shocks. In Indonesia, governments rely on administrative data, or deploy rapid surveys in communities to quantify the impacts of shocks. However, these approaches fail to provide insights at the desired spatial and temporal resolutions. Big (or alternative) data sources in combination with traditional data (e.g. surveys) provide a significant opportunity to plan and implement interventions in a targeted manner. The Government of Indonesia (GoI) has used big data sources in the past to monitor impacts of shocks. For instance GoI uses social media platform Twitter to nowcast in real time prices of key agricultural commodities, across markets all over Indonesia. However, since big data sources are originally not designed to sense impacts of shocks, a significant amount of preparatory analysis, in terms of what specific aspects of the end-user does the data capture, and how does this translate into insights, is required to assess its applicability in the context of shocks. Location and transaction-based information derived from transactions at ATM (Automatic Teller Machines) terminals offer a potentially interesting opportunity to sense economic impacts of shocks. Since \href{http://www.oecd.org/dev/asia-pacific/saeo-2019-Indonesia.pdf}{seventy-five percent of the Indonesian population currently have access to banking services}, this data source offers the potential to “sense” impacts resulting from shocks at high resolutions, across space and time. In this study we report the results from an initial assessment of location and transactions-based insights obtained from the ATM network of a major private sector bank in Indonesia, for the years 2014 and 2015. We perform preparatory analysis and assess the feasibility for use of this data to sense differential impacts of shocks across income groups at high spatial and temporal resolutions.
\section{Data and Methods}
The anonymised data set consists of individual customer transaction logs for years 2014 and 2015, performed at ATM terminals across the island of Java, by bank account holders of a major private sector bank in Indonesia. For each year, every log consists of an anonymised user identification number (ID) that is related to the unique bank account number, type (Cash Withdrawal or Others), and amount associated with each transaction. Additionally, yearly reported incomes for the corresponding unique identification number, as captured in the bank's Know Your Customer (KYC) records were also obtained. In the KYC records, yearly incomes are classified according to the following groups -  sd. 1 Juta, >1 Juta - 3 Juta, >3 Juta - 5 Juta, >5 Juta - 10 Juta, >10 Juta - 25 Juta, >25 Juta - 50 Juta, >50 Juta - 100 Juta and  >100 Juta, wherein Juta refers to a million Indonesian Rupiah (IDR). Since the proportion of customers within the income groups >3 Juta - 5 Juta (i.e. lower-middle income group; 158358 and 174848 unique customer IDs for years 2014 and 2015), sd. 1 Juta (i.e. low income group; 33918 and 31109 unique customer IDs for years 2014 and 2015) and  >100 Juta (high income group; 11362 and 15398 unique customer IDs for years 2014 and 2015) groups, was greater than 10\% across the entire customer base, the data set therefore for each year was subset to these specific income groups.

The data set also consists of a geotag for every ATM terminal. The geotag for each ATM terminal is derived from the corresponding nearest postal code. Eighty percent (9 million transactions for each year) of the transaction logs captured across both the years were tagged as “Cash Withdrawals” as transaction type, therefore logs with this transaction type were further filtered, and used for subsequent analysis. Since a customer transacts multiple times across multiple terminals all across the year, aggregate statistics of transactions from multiple cash withdrawals across all terminals, for each customer were derived separately for each year, and used as predictors. For each customer and for ~ 260000 customers in total, aggregated cash withdrawal transactions included per year sum, average, standard deviation and median values of cash withdrawal amounts. Additionally, as customers withdraw from multiple ATM terminals, total inter-terminal distance traversed by each customer in each year was calculated using the geotags of ATM terminals. Processing of predictors was implemented in R statistical environment (v 3.4.1) \cite{t13}, using the dplyr \cite{w17}, tidyr \cite{wh20} and data.table \cite{ds19} packages. Calculation of the total inter-terminal distance traversed, was calculated using the Euclidean distance measure (converted to kilometers) using the sp package \cite{pb05} in R statistical environment (v 3.4.1) \cite{t13}. Per customer, per year summary statistics of cash withdrawals (i.e. sum, median, standard deviation, average and total distance traversed) for the selected income groups (i.e. sd 1 juta or lower income group,  >3 juta - 5 juta or lower-middle income group and >100 juta or high income group) were visualized using box plots, that were generated using the ggplot2 \cite{w16} package in R statistical environment (v 3.4.1) \cite{t13}. 

For each customer, and in each year, cash withdrawal aggregates, and total inter-terminal distance traversed for that specific year were used as predictors, and the reported income group for the corresponding customer was used as the predictand, and the relationship between the two was modelled using Extreme Gradient Boosting (XGB) algorithm, using package xgboost \cite{chbktccmczlxlgl20} in R statistical environment (v 3.4.1) \cite{t13}. Hyperparameter tuning for the XGB model was performed using the mlr package \cite{blksrscj16} in R statistical environment (v 3.4.1) \cite{t13}. Tunable parameters which included the booster type, maximum depth, gamma, minimum child weight, subsample, colsample by tree, eta, error metric and evaluation metric, were tuned using a bracket of lower and upper limit values for each parameter. Package CARET \cite{k20} in R statistical environment (v 3.4.1) \cite{t13} was used to partition the data for training, validation and testing, construction of the confusion matrix, and to obtain the corresponding accuracy, specificity and sensitivity values. For comparison, a baseline model that was tuned to predict the most frequent customer income group, i.e. the lower-middle income group, was developed using the ZeroR function in OneR package \cite{j17} in R statistical environment (v 3.4.1) \cite{t13}. Accuracy, recall, precision and F1 scores of the XGB model were compared with that of the baseline model, to provide contextual understanding of the results. In order to aid interpretation of the XGB model, partial dependence plots were generated using the pdp package \cite{g17} in R statistical environment (v 3.4.1) \cite{t13}. The predictors used are known to be highly correlated, and since feature importance plots are unable to capture inherent correlations between the features, feature importance plots were produced based on SHAP values and their interactions, using the SHAPforxgboost package \cite{lj20} in R statistical environment (v 3.4.1) \cite{t13}. To further validate the findings of the model, a Kruskal-Wallis hypothesis test was performed to test for significant relationship between the top predictor obtained from the XGB model, and customer income group in R statistical environment (v 3.4.1) \cite{t13}.
\section{Results and Discussion}
\begin{figure}
  \centering
  \includegraphics[width=12cm]{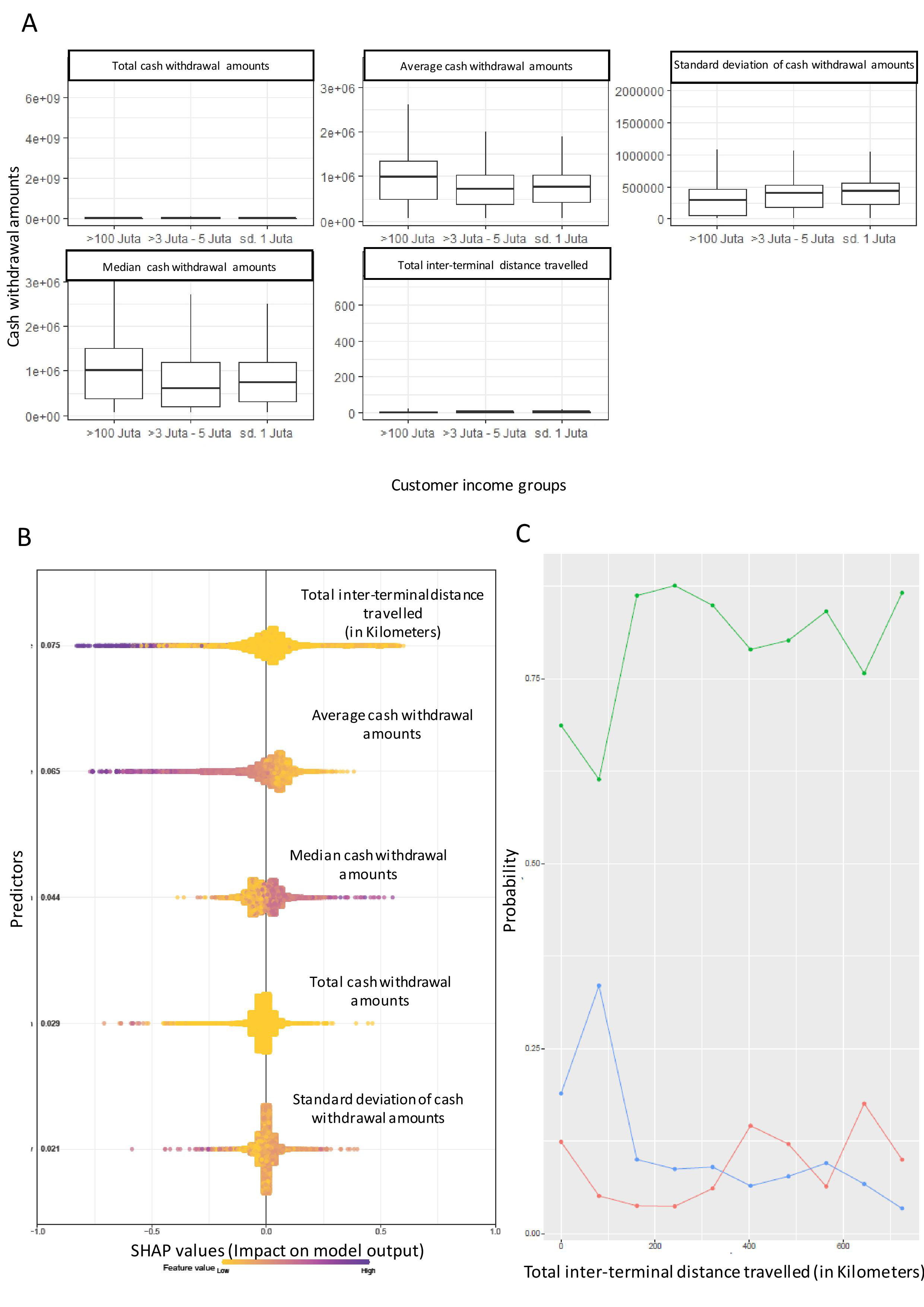}
  \caption{(A) Descriptive Statistics of transaction and location based characteristics obtained from Automatic teller machines (ATM's) in 2014, (B) Variable importance plots reflecting the SHAP values in 2014 for predictors used in Extreme gradient boosting, (C) Partial dependence plot of the total inter-terminal distance traversed by a customer for cash withdrawals in 2014}
  \label{fig:test}
\end{figure}

Descriptive statistics for 2014 in panel A in Figure \ref{fig:test}, show that yearly mean values of standard deviation per cash withdrawal, and the total inter-terminal distance traversed across multiple ATM terminals by low and lower-middle income group is relatively higher in comparison to high income group. The same was also observed in year 2015. 

Predictive modelling using XGB, wherein customer’s income group (obtained from the banks KYC records), was predicted using predictors derived from cash withdrawal transactions at ATM terminals, revealed that the customer income group can be predicted with 80\% accuracy, with relatively high recall values (0.53 for high income group, 0.79 for lower-middle and 0.59 for low income group for 2014/0.59 for high, 0.80 for lower-middle and 0.60 for low income group for 2015). However, the model yielded relatively low precision values for both the years (0.13 for high, 0.97 for lower-middle and 0.14 for low income group for 2014/0.10 for high, 0.98 for lower-middle and 0.05 for low income group for 2015). In contrast, the baseline model obtained a predictive accuracy of 78\% with null values for precision and recall for low and high income groups, for both the years. The XGB model therefore provided significantly better results in terms of predictive power in comparison to the baseline model. As seen in panel B in Figure \ref{fig:test}, feature importance plots based on SHAP values, obtained for each predictor, revealed that the per year total inter-terminal distance traversed across multiple ATM terminals by a customer was the most important predictor in 2014. The same was also observed in 2015. SHAP values further suggest that the resolving capabilities of total inter-terminal distance traversed across multiple ATM terminals is significantly higher for lower income groups (blue colored dots for this predictor to the left in panel B in Figure \ref{fig:test}). Partial dependence plot, as seen in panel C of Figure \ref{fig:test}, for total inter-terminal distance traversed across multiple ATM terminals in 2014, further confirms that the probability of predicting lower-middle class income group using the total per year per customer inter-terminal distance traversed, is significantly higher than low and high income groups. Kruskal-Wallis test further confirmed this finding, as the test identified a significant difference in total inter-terminal distance traversed per year between the customer income groups (Kruskal Wallis Chi-squared test, p-value < 2.2e-16), with significantly high median values for low-middle income group (7.21 kilometers in 2014 and 2015) in comparison to high (2.55 in 2014 and 0.67 kilometers in 2015) and low income group (6.47 kilometers in 2014 and 2015). 

These results reveal that customers in the lower-middle income group perform cash withdrawals across diverse ATM terminals consistently, in comparison to low and high income group customers. It has been previously shown that Indonesia's lower-middle class display relatively higher mobility, specifically in relation to income generation opportunities \cite{rhf20}, which is reflected by these results. Although there were no notable shocks in 2014 and 2015, one can extrapolate these findings to a shock event. For instance, relatively lower inter-terminal distances traversed specifically by the lower-middle income group, in comparison to the predicted counterfactual values, can help identify who is impacted and when. Although additional work is needed to analyse transaction logs with higher temporal frequencies and spatial resolutions in the event of a shock, the sensing capabilities have the potential to enable timely responses to shocks by governments and development organisations.  
\clearpage
\clearpage
\section{Broader Impact}
In this work we present evidence in relation to the opportunity offered by transaction and location based insights obtained from Automatic Teller Machines (ATM’s), to sense economic shocks and their impacts on populations in near real time, to specifically understand who is being impacted, where is the impact, and by how much. Although these insights are helpful in enabling governments and development agencies to respond to such shocks, it relies on a data source that is confidential and not collected or maintained by either the government or development agencies. This data belongs to the banks and credit rating agencies, and is used to understand the financial characteristics,and spending patterns of their customers. Banks and credit rating agencies are often interested in longer term effects of shocks on their customers, while governments and their partners are interested in relatively short to medium term impacts. Therefore it is necessary to balance the interests of the banks, and in response gain access to their data resources. For instance, new use cases from this data set, such as the relationship between customers' response to short term impacts to the long term ones, and the resulting changes in characteristics such as credit-worthiness, that can potentially help the bank, should be explored further. One particular challenge currently with the data source is customer income groups. Different banks have varying capacities in validating customer reported incomes, which will significantly impact the quality of the data and derived insights. Additionally there is an intriguing spatial bias in the methodology too, as we currently "assume" that customers transacting at a specific terminal also reside in the same location, especially since we don’t have access to the customers home location. In conclusion,    
this is the first time wherein ATM transaction data is being explored for use as a real time economic-shock sensing system in Indonesia, although the opportunities are galore in terms of enabling rapid response by government agencies and their partners, additional exploration will potentially reveal further shortcomings of the data, methods and the corresponding assumptions. 
\section{Acknowledgment}
The authors thank colleagues at the bank for their support in enabling access to transactions logs from their Automatic Teller Machines. The authors would also like to thank the contributions of team members at Pulse Lab Jakarta, specifically in relation to reviewing this work. Authors would like to thank the Department of Foreign Affairs and Trade of the Government of Australia, along with Ministry of National Development Planning (BAPPENAS) of the Government of Indonesia for providing consistent support and funding for this work. The authors would like to declare that there are no competing interests for this work.
\printbibliography
\end{document}